\documentclass[a4paper,12pt]{article}

\usepackage[T2A]{fontenc}
\usepackage[utf8]{inputenc}   
\usepackage[russian]{babel}

\usepackage{CJKutf8}

\usepackage{amsmath,amstext,amsfonts,amsthm,amssymb}
\usepackage{eucal}
\usepackage{graphicx}
\renewcommand{\subsubsection}[1]{\addtocounter{subsubsection}{1}
{\ \\[3pt]\bf \thesubsubsection. \  #1} }

\swapnumbers

{  \theoremstyle{definition}

}
\newcommand{\Alt}{\operatorname{Alt}}

\newcommand{\diag}{\operatorname{diag}}

\newcommand{\Fil}{\operatorname{Fil}}

\newcommand{\Ima}{\operatorname{Im}}

\newcommand{\Spec}{\operatorname{Spec}}

\newcommand{\hra}{\hookrightarrow}
\newcommand{\iso}{\overset{\sim}{\longrightarrow}}

\newcommand{\lra}{\longrightarrow}

\newcommand{\lla}{\longleftarrow}

\newcommand{\bea}{\begin{eqnarray*}}
\newcommand{\eea}{\end{eqnarray*}}
\newcommand{\bean}{\begin{eqnarray}}
\newcommand{\eean}{\end{eqnarray}}

\newcommand{\tK}{\tilde K}

\newcommand{\fg}{\mathfrak g}

\newcommand{\fm}{\mathfrak m}

\newcommand{\fn}{\mathfrak n}

\newcommand{\fsl}{\mathfrak{sl}}

\newcommand{\CA}{\mathcal{A}}

\newcommand{\CH}{\mathcal{H}}
\newcommand{\CI}{\mathcal{I}}

\newcommand{\CL}{\mathcal{L}}

\newcommand{\CO}{\mathcal{O}}

\newcommand{\BA}{\mathbb{A}}

\newcommand{\BF}{\mathbb{F}}

\newcommand{\BN}{\mathbb{N}}

\newcommand{\BQ}{\mathbb{Q}}

\newcommand{\BZ}{\mathbb{Z}}

\newcommand{\nc}{\newcommand}

\nc{\Id}{\text{Id}}
\nc{\la}{\lambda}

\begin{document}


\bigskip\bigskip

\centerline{\bf DE RHAM - WITT KZ EQUATIONS}

\

\centerline{\it Crystalline bozonization}

\

\centerline{Vadim Schechtman and Alexander Varchenko}

\


\bigskip\bigskip

\centerline{December 5, 2022}

\


\



\begin{CJK}{UTF8}{min}


\end{CJK}

\hskip 4 cm {\it Те, кому приходилось столкнуться с векторами Витта  

\hskip 4cm в собственной 
работе, часто относятся к ним 

\hskip 4cm с теплотой

\hskip 10cm Д. Каледин}\footnote{See [K]} 

\

\centerline{\bf Introduction}

\

In this note 
we propose a de Rham - Witt version of the derived KZ equations, and of their hypergeometric realizations, 
cf. [SV2]. In particular  
we describe a de Rham - Witt version of Theorems  $7.2.5'$, $7.2.5''$  from [SV1].

We also propose de Rham - Witt versions 
of some classical theorems related to arbitrary hyperplane arrangements.


\

{\it Contents of the paper}

\

In \S 1 we recall some background material about the de Rham - Witt complex. In \S 2 we present a de Rham - Witt version of the first part of [SV1] concerning arbitrary hyperplane arrangements over $\BF_p$. 
In 2.3.3 - 2.3.7 we discuss also a motivic version as suggested by A.Beilinson. In \S 3 we present 
a de Rham - Witt version of [SV2] which in its turn is a development of [SV1], Part 2. For simplicity 
of exposition  
we discuss only the case $\fg = \fsl_2$ but the generalization to arbitrary Kac - Moody algebras 
is straightforward.  

\

{\it Acknowledgements}

\ 

We are grateful to Alexander Beilinson for an interesting discussion and consultations; the idea of 2.3.3 - 2.3.7 belongs to him, and the proofs are obtained jointly with him. We express our gratitude to H\'el\`ene  Esnault for important corrections.

Our reflections on the topics of this paper 
have been triggered by a talk of Rub\'en Mu\~noz-Bertrand; we are thankful to him. 

\bigskip\bigskip

\

\centerline{\bf \S 1. The de Rham - Witt complex  } 

\

\

{\bf 1.1. De Rham - Witt complex of a polynomial ring.} Cf. [B], \S 2.

$p$ - a prime number. $W(\BF_p) = \BZ_p,\ F = \Id, V = p\Id$. Thus
$$
W_{n}(\BF_p) := W(\BF_p)/V^{n}W(\BF_p) = \BZ/p^n\BZ.
$$ 

(a) {\it Witt ring}\ 

Let $A := \BF_p[t_1,\ldots, t_n]$. 

We consider, together with S.Lubkin et al., cf. [L, I1, I2, B], a (nonnoetherian) algebra  
$$
C = \cup_{i\geq 0} \BQ_p[t_1^{p^{-i}}, \ldots, t_n^{p^{-i}}] 
$$

{\bf 1.1.1. Remark.} $C$ is filtered by subalgebras
$$
C_m = \cup_{0\leq i\leq m} \BQ_p[t_1^{p^{-i}}, \ldots, t_n^{p^{-i}}],
$$
and each
$$
C_m = \BQ_p[t_1^{p^{-m}}, \ldots, t_n^{p^{-m}}].
$$

We may also consider a bigger ring 
$$ 
\hat C = \{\sum_{(i_1,\ldots, i_n)\in \BN[p^{-1}]} a_{i_1,\ldots,i_n}t_1^{i_1}\ldots 
t_n^{i_n} \}
$$
(formal power series).

Both $C$ and $\hat C$ are equipped with a $\BQ_p$-linear automorphism $F$, 
$$
F(t_i) = t_i^p,
$$
and we define $V := pF^{-1}$.

\

{\it Notations.} For $i\in \BZ[p^{-1}]\subset \BQ\subset \BQ_p$, 
$$
d(i) := - v_p(i);
$$
for $I = (i_1,\ldots i_n)\in \BZ[p^{-1}]^n$, 
$$
|I| = \sum i_k,
$$
$$
d(I) = \max\{ d(i_k)\}.
$$

\

The Witt  ring $W(\BF_p[t_1, \ldots, t_n])$ is a subring of $\hat C$, namely
$$
W(\BF_p[t_1, \ldots, t_n]) = \{\sum_{(i_1,\ldots, i_n)\in \BN[p^{-1}]^n} a_{i_1,\ldots,i_n}t_1^{i_1}\ldots 
t_n^{i_n}|\ \forall I\ v_p(a_I)\geq d(I),\ \lim_{|I|\lra\infty} a_I = 0 \},
$$
cf. [L], VI (a). It is stable with respect to $F$ and $V$.

\

(b) {\it De Rham complex $\Omega^\bullet_{C/\BQ_p}$}

\

Consider the de Rham complex $\Omega^\bullet_{C/\BQ_p}$. 


Each $m$-form $\omega\in \Omega^m_{C/\BQ_p}$  may be uniquely written as 
$$
\omega = \sum_{1\leq i_1 < \ldots < i_m\leq n} a_{i_1,\ldots, i_m}(t)d\log t_{i_1}\wedge \ldots 
d\log t_{i_m}
$$
where $a_I(t)\in C$. 

Define  automorphisms $F:\ \Omega^m_{C/\BQ_p}\iso \Omega^m_{C/\BQ_p}$ by
$$
F(\sum_I a_{i_1,\ldots, i_m}(t_1,\ldots, t_n)d\log t_{i_1}\wedge \ldots 
d\log t_{i_m}) = 
$$
$$
\sum_I a_{i_1,\ldots, i_m}(t^p_1,\ldots, t^p_n)d\log t_{i_1}\wedge \ldots 
d\log t_{i_m}. 
$$

{\it Compatibility with $d$}:

We have
$$
dF = pFd
$$
As before, we put $V:= pF^{-1}$.

Thus we have two morphisms $F, W:\ \Omega^\bullet_{C/\BQ_p}\lra \Omega^\bullet_{C/\BQ_p}$ which satisfy:
$$
FV = VF = p,\ FdV = d,\ dF = pFd,\ Vd = pdV,\ 
$$
$$
V(x\cdot Fy) = Vx\cdot y
$$

\

(c) {\it De Rham - Witt complex}

\

Let us say that $\omega\in \Omega^m_{C/\BQ_p}$ is {\it integral} if all coefficients of all $a_I(t)$ belong to $\BZ_p$. 
Define subspaces
$$
E^m = \{\omega\in \Omega^m_{C/\BQ_p}|\ \omega, d\omega\text{ are integral}\},
$$
for example
$$
E^0 = \cup_{i\geq 0} p^i\BZ_p[t_1^{p^{-i}}, \ldots, t_n^{p^{-i}}] .
$$
They form a subcomplex $E^\bullet\subset \Omega^\bullet_{C/\BQ_p}$ stable with respect to $F$ and $V$. 

Define
$$
\Fil^aE^m = V^aE^m + d(V^aE^{m-1})\subset E^m;
$$
these subspaces form a subcomplex $\Fil^a E^\bullet\subset E^\bullet$,
and we may consider the quotient complex
$$
W_a\Omega^\bullet_A = E^\bullet/\Fil^aE^\bullet.
$$
When $a\geq 1$ varies, these complexes form a projective system.
$$
W\Omega^\bullet_A = \lim_{\lla} W_a\Omega^\bullet_A.
$$
\

{\bf 1.2. De Rham - Witt complex of Laurent polynomials.} The construction is similar, cf. [I2] II.2.

 Consider the case of one variable. 
 $$
 A = \BF_p[t, t^{-1}],
 $$
 $$
 C = \cup_{r\geq 0}\BQ_p[t^{1/p^r}, t^{- 1/p^r}]
 $$
$$
F:\ C\iso C,\ F(a(t, t^{-1})) = a(t^p, t^{-p}),\ V = pF^{-1}.
$$ 
The de Rham complex $\Omega^\bullet_{C/\BQ_p}$ has length $1$:
$$
0 \lra C = \Omega^0_{C/\BQ_p} \lra \Omega^1_{C/\BQ_p}\lra 0.
$$ 
Each form $\omega\in \Omega^1_{C/\BQ_p}$ can be uniquely written as 
$$
 \omega = \sum_{i\in \BN[p^{-1}]} a_i(t, t^{-1}) d\log t, \ a_i\in C, 
$$
$$
F:\ \Omega^1_{C/\BQ_p}\iso \Omega^1_{C/\BQ_p},\ 
$$
$$
F(\sum_{i\in \BN[p^{-1}]} a_i(t, t^{-1}) d\log t) = \sum_{i\in \BN[p^{-1}]} a_i(t^p, t^{-p}) d\log t,
$$
$$
V = pF^{-1}.
$$
Let us say that $\omega\in \Omega^m_{C/\BQ_p}$ is {\it integral} if all coefficients of all $a_I(t)$ belong to $\BZ_p$. 
Define subspaces
$$
E^m = \{\omega\in \Omega^m_{C/\BQ_p}|\ \omega, d\omega\text{ are integral}\},
$$
for example
$$
E^0 = \cup_{i\geq 0} p^i\BZ_p[t^{p^{-i}}, t^{- p^{-i}}] .
$$
They form a subcomplex $E^\bullet\subset \Omega^\bullet_{C/\BQ_p}$ stable with respect to $F$ and $V$. 

Define
$$
\Fil^aE^m = V^aE^m + d(V^aE^{m-1})\subset E^m;
$$
these subspaces form a subcomplex $\Fil^a E^\bullet\subset E^\bullet$,
and we may consider the quotient complex
$$
W_a\Omega^\bullet_A = E^\bullet/\Fil^aE^\bullet.
$$
When $a\geq 1$ varies, these complexes form a projective system.
$$
W\Omega^\bullet_A = \lim_{\lla} W_a\Omega^\bullet_A.
$$

\

{\bf 1.3. Shift.} Given $z\in\BF_p$, let
$$
A = \BF_p[t, (t - z)^{-1}] = \BF_p[t - z, (t - z)^{-1}].
$$
We introduce an algebra
$$
C = \cup_{r\geq 0}\BQ_p[(t - z)^{1/p^r}, (t - z)^{- 1/p^r}],
$$
an isomorphism
$F:\ C\iso C$, 
$$
\ F(a(t - z, (t - z)^{-1})) = a((t - z)^p, (t - z)^{-p}),\ 
$$
and an endomorphism $V: C\lra C$,
$$
V = pF^{-1}.
$$
Then we proceed as before: $1$-forms will have the form
$$
 \omega = \sum_{i\in \BN[p^{-1}]} a_i(t - z, (t - z)^{-1}) d\log (t - z),  
$$
etc.

\

{\bf 1.4. Laurent polynomials of several variables.} Let $I\subset [n] = \{1, \ldots, n\}$, 
and 
$$
A = \BF_p[t_1,\ldots, t_n, t_i^{-1}, i\in I]
$$
We proceed as before, starting with an algebra 
$$
C = \cup_{r\geq 0}\BQ_p[t_1^{1/p^r},\ldots, t_n^{1/p^r}, t_i^{\pm 1/p^r}, i\in I],
$$
etc.

\

The multiplicative group $A^*$ is a direct sum of $\BF_p^*$ and of a free abelian group generated by $t_i, \ i\in I$. The Teichm\"uller 
homomorphism
$$
T:\ A^*\lra W(A)^*\subset \hat C^* 
$$
sends $t_i$ to $t_i\in \hat C^*$. 

\

{\bf 1.5.} More generally, we will be interested in the de Rham - Witt complex of the ring
$$
A = \BF_p[t, z_1,\ldots, z_n, (t - z_1)^{-1},\ldots (t - z_n)^{-1}].
$$
Similarly to the previous considerations, we introduce a ring 
$$
C = \cup_{r\geq 0} \BQ_p[t^{1/p^r},  (t - z_1)^{\pm 1/p^r},\ldots (t - z_n)^{\pm 1/p^r}]
$$
and its completion $\hat C$. 

The complex $W\Omega^\bullet_A$ will be a subcomplex of $\Omega^\bullet_{\hat C/\BQ_p}$.

\

\newpage

\centerline{\bf \S 2. Du c\^ot\'e de chez Orlik - Solomon}

\

{\bf 2.1. Orlik - Solomon subalgebra, aka fixed points of Frobenius. }

\

{\bf Notation.} Given a whatever object $?$ acted upon by a Frobenius $F$, the superscript $?^{F-1}$ will denote 
$\{ x\in ?|\ Fx = x\}$.

\

{\bf 2.1.1.} Let $A = \BF_p[t, t^{-1}]$. Consider the fixed submodule of Frobenius
$$
W\Omega^{\bullet F - 1}_A\hra W\Omega^{\bullet}_A.
\eqno{(2.1.1)}
$$
We have 
$$
W\Omega^{0 F - 1}_A = \BZ_p,\ W\Omega^{1 F - 1}_A = \BZ_p d\log t,
$$
the differential in $\Omega^{\bullet F - 1}_A$ being zero.

The embedding (2.1.1) is a quasiisomorphism.

{\bf 2.1.2.} A more general case of 1.4 is similar: for
$$
A = \BF_p[t_1,\ldots, t_n, t_i^{-1}, i\in I]
$$
the subcomplex $W\Omega^{\bullet F - 1}_A$ is a free exterior $\BZ_p$-algebra with generators 
$d\log t_i, i\in I$ and zero differential,
$$
W\Omega^{\bullet F - 1}_A = \Lambda^\bullet_{\BZ_p} (\oplus_{i\in I} \BZ_p d\log t_i).
$$ 
It will be called the {\it Orlik - Solomon subalgebra}, 
and denoted also $W\Omega^{\bullet OS}_A$. 

The embedding
$$
W\Omega^{\bullet OS}_A\hra W\Omega^{\bullet}_A.
\eqno{(2.1.2)}
$$
is a quasiisomorphism. 

{\bf 2.2.} This is a general phenomenon, valid for any hyperplane arrangement. 
Namely, consider an affine space 
$$
\BA^n_{\BF_p} = \Spec \BF_p[t_1,\ldots, t_n]
$$
and a finite collection $\CH = \{H_i\}_{i\in I}$ of affine hyperplanes in it, $H_i$ being given 
by an equation $f_i(t) = 0$. 

Let
$$
U = \BA^n_{\BF_p}\setminus \bigcup_{i\in I} H_i = \Spec A 
$$
where
$$
A = \BF_p[t_1,\ldots, t_n, f_i^{-1},\ i\in I].
$$
We will use both notations $W\Omega^\bullet_A = W\Omega^\bullet(U)$. 

\

{\bf 2.2.1. Fixed points of Frobenius are closed.} 
If $x\in W\Omega^i_A$, $x = Fx$ then 
$$
dx = dFx = pFdx,
$$
i.e.
$$
(1 - pF)dx = 0.
$$
But the operator $1 - pF$ is invertible:
$$
(1 - pF)^{-1} = \sum_{i\geq 0} p^iF^i, 
$$
and the series in the RHS converges. Whence $dx = 0$.

We've got an embedding of complexes
$$
os: 
W\Omega^{\bullet OS}_A := W\Omega^{\bullet F - 1}_A \hra W\Omega^{\bullet}_A.
\eqno{(2.2.1)}
$$
where the first one is equipped with the zero differential.

\

{\bf 2.2.2. Logarithmic forms. } 
Let $H\subset \BA = \BA^n_{\BF_p}$ be a hyperplane given by an equation $f(t) = 0$. Let us choose new coordinates $u_1, \ldots, u_n$ in  
$\BA$ such that in new coordinates $H$ will be given by the equation $u_n = 0$. Thus 
we get an isomorphism
$$
\phi:\ \BA\setminus H\iso \Spec \BF_p[u_1, \ldots, u_{n-1}, u_n^{\pm 1}]
$$
We may apply 1.4 to the target and get closed $1$-forms 
$$
d\log u_n\in W\Omega^1(\BF_p[u_1, \ldots, u_{n-1}, u_n^{\pm 1}]),
$$
whence 
$$
d\log f := \phi^*d\log u_n\in W\Omega^1(\BA\setminus H).
$$
Otherwise one can define $d\log f$ as in [I2] 3.23, as 
$$
d\log f = d\tilde f/\tilde f
$$
where $\tilde f\in W\CO(\BA\setminus H)$ is the Teichm\"uller representative of $f$. 
 
\

{\bf 2.2.3.} Returning to our arrangement, we obtain  
closed $1$-forms $d\log f_i\in W\Omega^1_A$; by construction they belong to $W\Omega^{1 F - 1}_A$.

Compare with [I2] CH. I, (3.23). 

According to {\it op. cit.} I, Thm. 5.7.2
$$
W\Omega^{\bullet F - 1}_U\subset W\Omega^{\bullet}_U
$$
is the same as the subsheaf generated by logarithmic forms. 

\

{\bf 2.3. Orlik - Solomon.} Recall that the Orlik - Solomon algebra $\CA(\CH)$, cf. [OS], \S 2, is an 
$\BN$-graded  skew commmmutative ring with $\CA(\CH)_0 = \BZ$, the quotient of the exterior algebra with generators $(H_i), i\in I$, of degree 1, modulo the ideal   
generated by the elements
$$
(H_1,\ldots, H_p) := H_1\wedge\ldots\wedge H_p
$$
if $H_i$ are not in general position, and 
$$
\sum_{i=1}^{p+1} (-1)^i (H_1,\ldots,\hat H_i,\ldots, H_{p+1}) = 0
$$
if $H_1, \ldots, H_{p+1}$ are not in general position and $H_1\cap\ldots\cap H_{p+1}\neq \emptyset$, cf. [SV1], 1.2. 

Multiplication is given by concatenation. 

Recall that all $\CA(\CH)_p$ are free abelian groups. 

{\bf 2.3.1. Proposition.} {\it There exists a unique algebra map 
$$
\psi:\ \CA(\CH)\lra W\Omega^{\bullet OS}_A = W\Omega^{\bullet OS}(U)
$$
such that $\psi(H_i) = d\log f_i$.} $\square$

\

{\it Integer structure}

\

Let 
$$
W\Omega^{\bullet OS}(U)_\BZ = \Ima(\psi)\subset W\Omega^{\bullet OS}(U)
$$
denote the $\BZ$-subalgebra generated by $d\log f_i$.

\

{\bf 2.3.2. Proposition.} {\it   The map $\psi$ induces an isomorphism
$$
\psi:\ \CA(\CH)\iso W\Omega^{\bullet OS}(U)_\BZ,
$$
whence an isomorphism
$$
\psi_{\BZ_p}:\ \CA(\CH)\otimes\BZ_p \iso W\Omega^{\bullet OS}(U).
$$}

{\bf Proof} the same as in [OS], Thm. 5.2. $\square$
\


\

{\it Motivic variation:  Milnor $K$-theory and  Orlik - Solomon}

\

{\bf 2.3.3. } As was suggested by Beilinson, we may consider a multiplicative version of the above.

At the beginning we may replace $\BF_p$ by an arbitrary base field $k$. 

Let 
$$
k(A) = k(U) = k(\BA^n_k) = k(t_1, \ldots, t_n)
$$
be the field of rational functions on $\BA^n_k$. Let $K_\bullet^M(k(U))$ be the algebra of Milnor $K$-functors. 
Thus
$$
K_0^M(k(U)) = \BZ, \ K_1^M(k(U)) = k(t_1, \ldots, t_n)^*,
$$
and $K_\bullet^M(k(U))$ is a an $\BN$-graded skew commutative algebra 
generated by $K_{1}^M(k(U))$ subject to the Steinberg relation
$$
\{ x, 1 - x\} = 0,\ x, 1 - x\in  K_{1}^M(k(U)), 
\eqno{(2.3.1)} 
$$
cf. [M].  

Let
$$
\tK_1^M(U)\subset K_1^M(k(U))
$$
be the subgroup generated by $f_i$; it is a free abelian group with a base $\{ f_i, i\in I\}$. Let 
$$
\tK_\bullet^M(U)\subset K_\bullet^M(k(U))
$$
be the $\BZ$-subalgebra generated by $\tK_1^M(U)$. 

We have a map 
$$
\chi_{\BZ 1}:\ \tK_1^M(U)\lra \CA_1(\CH)
$$
sending $f_i$ to $(H_i)$. 

{\bf 2.3.3.1. Claim.} {\it The map $\chi_{\BZ 1}$  extends to 
an algebra map
$$
\chi_\BZ:\ \tK_\bullet^M(U)\lra \CA(\CH)
\eqno{(2.3.2)}
$$}

{\bf 2.3.3.2. Example. Identity "ch\`ere \`a Gelfand": magic of the Steinberg relation.} Consider an arrangement $\CH$ in $\BA^2_k = \Spec k[x, y]$ consisting 
of $3$ lines 
$$
\ell_1:\ x = 0,\ \ell_2:\   y = 0,\ \ell_3:\  x = y,
$$ 
so 
$$
U = \BA^2_k\setminus \bigcup_{i=1}^3 \ell_i.
$$  

We have a familiar relation
$$
\frac{1}{x(x - y)} - \frac{1}{y(x - y)} = - \frac{1}{xy}, 
$$
and its counterpart in $\CA(\CH)$. Let us write down its counterpart in $K_2^M(U)$. 

We have the following equalities in $K_2^M(k(U))\otimes \BZ[1/2]$:
$$
\{ x, x - y\} = \{x , x\} + \{ x, 1 - x^{-1}y\} = \{ x, 1 - x^{-1}y\};
$$
$$
\{ y, x - y\} = \{y, xy^{-1} - 1\} = \{y, xy^{-1}\} + \{y, 1 - x^{-1}y\} =    
$$
$$
= \{y, x\} + \{y, y^{-1}\} + \{y, 1 - x^{-1}y\} = \{y, x\}  + \{y, 1 - x^{-1}y\}.
$$
Here we have used the equality $\{ y, y\} = 0$; by that reason we had to 
tensorise by $\BZ[1/2]$.

On the other hand
$$
\{ x, 1 - x^{-1}y\} = \{ x, 1 - x^{-1}y\} + \{x^{-1}y, 1 - x^{-1}y\} = 
\{x x^{-1}y, 1 - x^{-1}y\} = \{ y, 1 - x^{-1}y\},  
$$
whence 
$$
\{ x, x - y\} - \{ y, x - y\} = - \{y, x\} = \{x, y\}.
$$
This is the Gelfand identity in $K_2^M$. $\square$

\

{\bf 2.3.4. Conjecture} (A.Beilinson) . {\it The map $\chi_\BZ$}  (2.3.2) {\it is an isomorphism. }

\

The map $\chi_\BZ$ composed with $\psi$ (see Prop. 2.3.2) gives rise to a map
$$
d\log:\ \tK^M_\bullet(U)\lra 
W\Omega^{\bullet OS}(U)_\BZ
\eqno{(2.3.4)}
$$
sending $f_i$ to $d\log f_i$.

\

{\bf 2.3.5. Motivic Orlik - Solomon.} {\it 
A canonical map 
$$
H^\bullet_{Mot}(U;\BQ)^{\diag}  \lra H^\bullet_{Mot}(U;\BQ)
$$
is an isomorphism. }

Here $H^\bullet_{Mot}(U; \BQ)$ denotes the motivic cohomology realized as appropriate eigenspaces of Adams operations 
in the Quillen $K$-theory, see [Be], 2.2, and $H^\bullet_{Mot}(U;\BQ)^{\diag}$ is its part consisting 
of all elements of type $(i,i)$.  

{\bf Proof.}   Our assertion says that all elements of $H^\bullet_{Mot}(U;\BQ)$ 
are of type $(i,i)$. This may be proven by induction on $n$. 
$\square$

\

{\bf 2.3.6.} Instead of rational motivic cohomology we could use more refined version, the Voevodsky's motivic 
cohomology, [MVW].  

It is also probable that a reduced version from [ES] works for $U$. 

\

Let us return to the de Rham - Witt realization.

\ 

{\bf 2.4. Theorem.} 
{\it The embedding} $os$ (2.2.1) {\it  is a quasiisomorphism.}

{\bf Proof} the same as in [OS], Thm. 5.2 which uses [Br] Lemme 5. Brieskorn uses the Lefschetz theorems 
on hyperplane sections which hold true for the crystalline cohomology. 
$\square$

\

{\bf 2.5. Turning on a connection, "Aomoto conjecture".} Given a collection of exponents 
$a_i\in \BZ_p,\ i\in I$, we get a closed one-form
$$
\omega = \sum_i a_id\log f_i\in W\Omega^{1 OS}_A,
$$
and a connection
$$
\nabla = d + \omega\wedge\bullet : W\Omega_A^i\lra W\Omega_A^{i+1}. 
$$
which restricted to $W\Omega^{\bullet OS}_A$ is simply the multiplication by $\omega$.

\

{\bf 2.6. Theorem} (Aomoto conjecture, a weak form). 
{\it  For sufficiently small $a_i$ (in the $p$-adic metric) 
the embedding
$$
(W\Omega^{\bullet OS}_A, \omega) \hra (W\Omega^{\bullet }_A, d + \omega)
$$
is a quasiisomorphism.}

{\bf Proof.} The same as in [SV1] Thm 4.6. $\square$

\

{\bf 2.7.} It is probable that a strong form of the Aomoto conjecture similar to [ESV] is valid as well. 
 
\


\centerline{\bf \S 3. De Rham - Witt Coulomb - KZ - Chevalley cocycle}

\

{\bf 3.1.} Consider a simple example. Suppose that $p > 2$. Suppose we are given $n$ numbers $a_i\in \BZ_p$, $1\leq i \leq n$. Consider an affine space
$$
\BA = \Spec\BF_p[t, z_1, \ldots , z_n]
$$
and a hyperplane arrangement in it consisting of hyperplanes $H_i: t = z_i,\ H_{ij}: z_i = z_j$. 

Consider the complement
$$
U = \BA\setminus ((\bigcup_i H_i)\cup (\bigcup_{i,j} H_{ij})) = \Spec\BF_p[t, z_1, \ldots , z_n, (t - z_i)^{-1}, (z_i - z_j)^{-1}]
$$
We have a Coulomb $1$-form
$$
\omega = - \sum_i 2a_id\log(t - z_i) + \sum_{i , j} a_ia_jd\log(z_i - z_j)\in W\Omega^1(U)
$$

\

{\it Construction of a hypergeometric cocycle}

\

We follow the pattern of [SV2], Section 2.

\ 

{\bf 3.2. Coulomb connection. }  Suppose we are given natural $n, N \geq 1$, and  an $n$-tuple 
$$
\fm = (m_1, \ldots, m_n)\in \BZ_p^n.
$$  
Let
$$
U_n = \Spec \BF_p[t_1,\ldots, t_N, z_1,\ldots z_n, (z_b - z_c)^{-1}], 
$$ 
$$
U_{n,N} =  \Spec A, \ A = \BF_p[t_1,\ldots, t_N, z_1,\ldots z_n, (t_i - t_j)^{-1}, 
(t_i - z_b)^{-1}, (z_b - z_c)^{-1}].
$$
We define the de Rham - Witt Coulomb closed $1$-form
$$
\omega_\fm = \sum_{1\leq b < c\leq n} \frac{m_bm_c}{2}d\log(z_b - z_c) + \sum_{1\leq i < j \leq N} 2
d\log(t_i - t_j) - 
$$
$$
- \sum_{1\leq i\leq N, 1\leq b\leq n} m_b d\log(t_i - z_b)\in W\Omega^1(A) = W\Omega^1(U_{n,N}).
$$
Given an invertible $\kappa\in\BZ_p$, we define a Coulomb  connection 
$$
\nabla_{Coul, KZ} = d + \frac{1}{\kappa}\omega_\fm:
\ W\Omega^\bullet(U_{n,N}) \lra W\Omega^{\bullet + 1}(U_{n,N}).
$$

\

{\bf 3.3. Chevalley complex.} Let $\fg$ be the Lie algebra $\fsl_2$ over $\BZ_p$ with generators $e, f, g$, $\fn = \BZ_p f\subset \fg$. 

For $m\in \BZ_p$ let
$$
M(m) = \oplus_{i\geq 0} \BZ_p f^iv
$$
be the Verma module, $ev = 0$; let  
$$
M(\fm) = M(m_1)\otimes_{\BZ_p}\ldots \otimes_{\BZ_p} M(m_n) = \oplus_{i\geq 0} M(\fm)_i
$$
We consider a Chevalley complex 
$$
C^\bullet(M(\fm)):\  0\lra M(\fm)\overset{d}\lra \fn\otimes_{\BZ_p} M(\fm),\ d(x) = f\otimes ex
$$
living in degrees $0, 1$, 
$$
C^\bullet(M(\fm)) = \oplus_{i\geq 0} C^\bullet(M(\fm))_i
$$
Let
$$
\Omega = \frac{1}{2}h\otimes h + e\otimes e + f\otimes f\in \fg\otimes \fg
$$
be the Casimir element.  For $1\leq i < j \leq n$ we will denote by 
$$
\Omega_{ij}:\ M(\fm) \lra M(\fm)
$$
the action of $\Omega$ through $i$-th and $j$-th factors. 

{\bf 3.4. Coulomb - KZ - Chevalley complex.} Let

We define a bicomplex 
$$
C^{\bullet\bullet}_{\fm, N} = \{W\Omega^p(U_{n,N})\otimes_{\BZ_p} C^q(M(\fm))_N\}_{ p\geq 0, 0\leq q\leq 1}.
$$
The horizontal differential
$$
d' = \nabla_{Coul, KZ} = d + \frac{1}{\kappa}\omega_\fm - \frac{1}{\kappa}\omega_{KZ}: 
$$
$$
\ W\Omega^\bullet(U_{n,N})\otimes_{\BZ_p} C^\bullet(M(\fm))_N \lra 
W\Omega^{\bullet + 1}(U_{n,N})\otimes_{\BZ_p} C^\bullet(M(\fm))_N
$$
acts on index $p$, 
where
$$
\omega_{KZ} := \sum_{1\leq b < c\leq n} \Omega_{bc} d\log(z_b - z_c): \ 
C^{pq}_{\fm,N} \lra  C^{p+1,q}_{\fm,N},
$$
the Casimirs $\Omega_{bc}$ acting through $M(\fm)$ factor. 

The vertical differential $d''$ acts on index $q$ and is induced by the differential 
$d_{Ch}$ in the Chevalley complex $C^\bullet(M(\fm))_N$. 

\

{\bf 3.5. Hypergeometric forms.}  They are defined by the same formulas as in [SV2] 2.2 but now they are considered 
as elements of the de Rham - Witt complex. 

For $b = (b_1, \ldots, b_n)\in \BN^n, |b| := \sum b_i = N$, define
$$
u_b = d\log(t_1 - z_1)\wedge\ldots\wedge d\log(t_{b_1} - z_1) + 
$$
$$
+ d\log(t_{b_1+1} - z_2)\wedge\ldots\wedge d\log(t_{b_1+b_2} - z_2) + 
$$
$$
\ldots + d\log(t_{b_1+\ldots + b_{n-1} +1} - z_n)\wedge\ldots\wedge d\log(t_{N} - z_n)\in 
W\Omega^N(U_{n,N}),
$$
and for $c = (c_1, \ldots, c_n)\in \BN^n, |c| = N - 1$, define
$$
u_c = - \kappa\biggl(d\log(t_2 - z_1)\wedge\ldots\wedge d\log(t_{c_1+1} - z_1) + 
$$
$$
+ d\log(t_{c_1+2} - z_2)\wedge\ldots\wedge d\log(t_{c_1+c_2+1} - z_2) + 
$$
$$
\ldots + d\log(t_{c_1+\ldots + c_{n-1} + 2} - z_n)\wedge\ldots\wedge d\log(t_{1} - z_n)\biggr)\in 
W\Omega^{N-1}(U_{n,N}),
$$

\

{\it Symmetrization}

\

Starting from here we suppose that $p > N$.

\


\ 

The symmetric group acts on $W\Omega^\bullet(U_{n,N})$ by permuting $t_i$.
For $w(t_1,\ldots,t_N)\in W\Omega^\bullet(U_{n,N})$ we denote 
$$
\Alt w(t_1,\ldots,t_N) = \sum_{\sigma\in\Sigma_N} (-1)^{|\sigma|}w(t_{\sigma(1)},\ldots, t_{\sigma(N)}). 
$$
We define
$$
w_b = \frac{1}{b_1!\ldots b_n!}\Alt u_b,\ w_c = \frac{1}{c_1!\ldots c_n!}\Alt u_c.
$$

{\bf 3.6.} Out of these forms we define a cochain in our bicomplex of total degree $N$: set 
$$
\CI_0 = \sum_{|b|=N} w_b\otimes f^{b}v\in C^{N,0}_{\fm,N} 
$$
where
$$
f^{b}v = f^{b_1}\ldots f^{b_n}v,
$$
and 
$$
\CI_1 = \sum_{|c|=N-1} w_c\otimes f\otimes f^{c}v\in C^{N-1,1}_{\fm,N}. 
$$

\

{\bf 3.7. Theorem.} {\it $\CI = (\CI_0, \CI_1)$ is a cocycle, i.e.  
$$
\nabla_{KZ,Coul}\CI_0 = 0,
$$
$$
d_{Ch}\CI_0 + \nabla_{KZ,Coul}\CI_1 = 0.
$$ }

{\bf Proof} the same as in [SV2], 2.5. $\square$

\

{\bf 3.8. Bozonization map.} Similarly to [SV2], 2.6 we can interpret our cocycle as a 
map of (filtered) complexes.

Namely, our bicomplex may be rewritten as a tensor product
$$
C^{\bullet\bullet}_{\fm, N} = W\Omega^\bullet(C^\bullet(M(\fm))_N)\otimes_{W\Omega^\bullet(U_n)} 
W\Omega_\fm^\bullet(U_{n,N})
$$
where
$$
W\Omega^\bullet(C^\bullet(M(\fm))_N) = W\Omega^\bullet(U_n)\otimes_{\BZ_p}C^\bullet(M(\fm))_N 
$$
equipped with the KZ connection
$$
\nabla_{KZ} = d_z - \frac{1}{\kappa}\omega_{KZ}, 
$$
and
$$
W\Omega_\fm^\bullet(U_{n,N}) = W\Omega^\bullet(U_{n,N})
$$
equipped with the Coulomb connection
$$
\nabla_{Coul} = d_z + \frac{1}{\kappa}\omega_{Coul}. 
$$  
Define the dual Chevalley complex 
$$
C_\bullet(M(\fm)^*)_N = (C^\bullet(M(\fm))_N)^*
$$
living in degrees $-1, 0$. The $N$-cocycle $\CI$ gives rise to 
a map of complexes 
$$
\eta(\CI):\ W\Omega^\bullet(C_\bullet(M(\fm)^*)_N)\lra W\Omega_\fm^\bullet(U_{n,N})^{\Alt}[N].
\eqno{(3.8.1)}
$$
Moreover, it is a map of filtered complexes, where 
$$
F^iW\Omega^\bullet(C_\bullet(M(\fm)^*)_N)\subset W\Omega^\bullet(C_\bullet(M(\fm)^*)_N)
$$
is the subcomplex of degree $\geq i$ forms, and
$$
F^iW\Omega_\fm^\bullet(U_{n,N})\subset W\Omega_\fm^\bullet(U_{n,N})
$$
is the subcomplex of forms having $\geq i$ of differentials $dz_a$.   

Let $\CL(\fm, N)$ denote the structure sheaf $\CO_{U_{n,N}})$ equipped with the Coulomb de Rham - Witt connection
$$
\nabla_\fm = d + \frac{1}{\kappa}\omega_\fm.
$$
Let $p: U_{n,N} \lra U_n$ be the projection. The direct image $Rp_*\CL(\fm, N)$ carries the Gauss - Manin 
connection $\nabla_{GM}$, the de Rham - Witt version of [KO].   

The induced by (3.8.1) map of $E_1$-pages of spectral sequences will be 
$$
\eta^i:\ (W\Omega^\bullet(H_i(M(\fm)^*)_N)(U_n), \nabla_{KZ}) \lra
$$
$$ 
\lra (W\Omega^\bullet(Rp_*^{N-i}(\CL(\fm, N)^{\Alt})(U_n), \nabla_{GM}),\ i = 0, 1.
$$

\


{\bf 3.9. Theorem.} {\it The maps $\eta^i$ are isomorphisms for $1/\kappa$ close to zero 
in the $p$-adic metrics.}

{\bf Proof.} This is a corollary of Thm. 2.6.  $\square$

\

{\bf 3.10.} All the above generalizes to the case of an arbitrary symmetrizable Kac - Moody algebra $\fg$. 
The details are left to an interested reader.

\

\

\

\

\centerline{\bf References}

\

[Be] A.Beilinson,  Higher regulators and values of $L$-functions, {\it Current problems in mathematics} {\bf 24}, 181–238,
{\it Itogi Nauki i Tekhniki, VINITI}, Moscow, 1984. 



[B] S.Bloch, Crystals and de Rham - Witt connections, {\it J. Inst. Math. Jussieu} {\bf 3} (2004), 315 - 326.

[Br] E.Brieskorn, Sur les groupes de tresses (d'apr\`es V.I.Arnold), S\'eminaire Bourbaki {\bf 401}, 1973.

[ES] J.N.Eberhardt, J.Scholbach, Integral motivic sheaves and geometric representation theory

[ESV] H.Esnault, V.Schechtman, E.Viehweg, Cohomology of local systems on the complement of hyperplanes, 
{\it Inv. Math.} {\bf 109} (1992), 557 - 561. 

[I1] L.Illusie, Complexe de de Rham - Witt,  {\it Journées de Géométrie Algébrique de Rennes}, 83–112, {\it Ast\'erisque} {\bf 63}, 1979. 

[I2] L.Illusie, Complexe de de Rham - Witt et cohomologie cristalline, {\it Ann. Sci. ENS} {\bf 12} (1979), 
501 - 661.

[K] D.Kaledin, Witt vectors, commutative and noncommutative, Russian Math. Surveys, {\bf 73} (2018), 3 - 34.

[KO] N.Katz, T.Oda, On the differentiation of De Rham cohomology classes with respect to parameters, 
{\it J. Math. Kyoto Univ.} {\bf 8} (1968), 199 - 213. 



[L] S.Lubkin, Generalization of $p$-adic cohomology; bounded Witt vectors. A canonical lifting 
of a variety in characteristic $p\neq 0$ back to characteristic zero, {\it Compos. Math.} {\bf 34} 
(1977), 225 - 277. 

[MVW] C.Mazza, V.Voevodsky, Ch.Weibel, Lecture notes on motivic cohomology

[M] J.Milnor, Introduction to algebraic $K$-theory, {\it Annals of Math. Studies} {\bf 72}, Princeton University Press, Princeton, NJ.; University of Tokyo Press, Tokyo, 1971.

[OS] P.Orlik, L.Solomon, Combinatorics and topology of complements of hyperplanes, {\it Inv. Math.} 
{\bf 56} (1980), 167 - 189.

[SV1] V.Schechtman, A.Varchenko, Arrangements of hyperplanes and Lie algebra homology, {\it Inv. Math.} 
{\bf 106} (1991), 139 - 194. 

[SV2] V.Schechtman, A.Varchenko, Derived KZ equations, {\it  J. Singul.} {\bf 25} (2022), 422 – 442. 


\

V.S.: Institut de Math\'ematiques de Toulouse, Universit\'e Paul Sabatier, 118 route de Narbonne, 31062 Toulouse, France;\ Kavli IPMU, 5-1-5 Kashiwa-no-ha, Kashiwa-shi, Chiba, 277-8583 Japan

\

A.V.: Department of Mathematics, University of North Carolina at Chapel Hill,
Chapel Hill, NC 27599-3250, USA;\ Faculty of Mathematics and Mechanics, Lomonosov
Moscow State University, Leninskiye Gory 1, 119991 Moscow GSP-1, Russia

\

\

\end{document}